\documentclass{moriond}

\usepackage{amssymb}

\def\be{\begin{equation}}
\def\ee{\end{equation}}
\def\bea{\begin{eqnarray}}
\def\eea{\end{eqnarray}}

\newcommand{\Photo}{}

\begin{document}
\vspace*{4cm}
\title{Scalar field dynamics around black holes: \\ superradiant instabilities and binary evolution}

\author{Giuseppe Ficarra}

\address{Department of Physics, King's College London, Strand, London, WC2R 2LS, United Kingdom}

\maketitle

\abstract{In classical general relativity astrophysical black holes can be affected by  the superradiant instability when gravity is minimally coupled to a light bosonic field. The majority of phenomenological studies have focused on the idealized case in which the black hole is initially surrounded by a single mode superradiant seed. By studying the evolution of a scalar field with multiple modes initial data in a quasiadiabatic approximation, we show that the dynamics is more involved and depend on the initial seed energy and the amplitude ratio between the modes. We also present preliminary results of the dynamical evolution of a massive scalar field around a Newtonian and a fully relativistic, black hole binary.}

\section{Introduction}
The interaction between black holes and ultralight bosonic fields, that represent popular dark matter candidates or axion-like particles, has opened new avenues to probe for beyond-standard model physics. In particular, Kerr black holes can undergo a superradiant instability\cite{Brito:2015oca} which actually turn astrophysical black holes into particle detectors \cite{Arvanitaki:2009fg,Arvanitaki:2010sy,Brito:2014wla}. Moreover the presence of a scalar field around merging black hole binaries could induce modifications in the gravitational waveform and studies in this direction have been recently started\cite{Baumann:2018vus,Baumann:2019ztm,Berti:2019wnn,Wong:2019kru,Wong:2020qom,Ikeda:2020xvt}. We use geometric units $G = c =1$.

\section{Superradiant instabilities and multiple modes}
\begin{figure}[ht!]
    \centering
    \includegraphics[width=.425\textwidth]{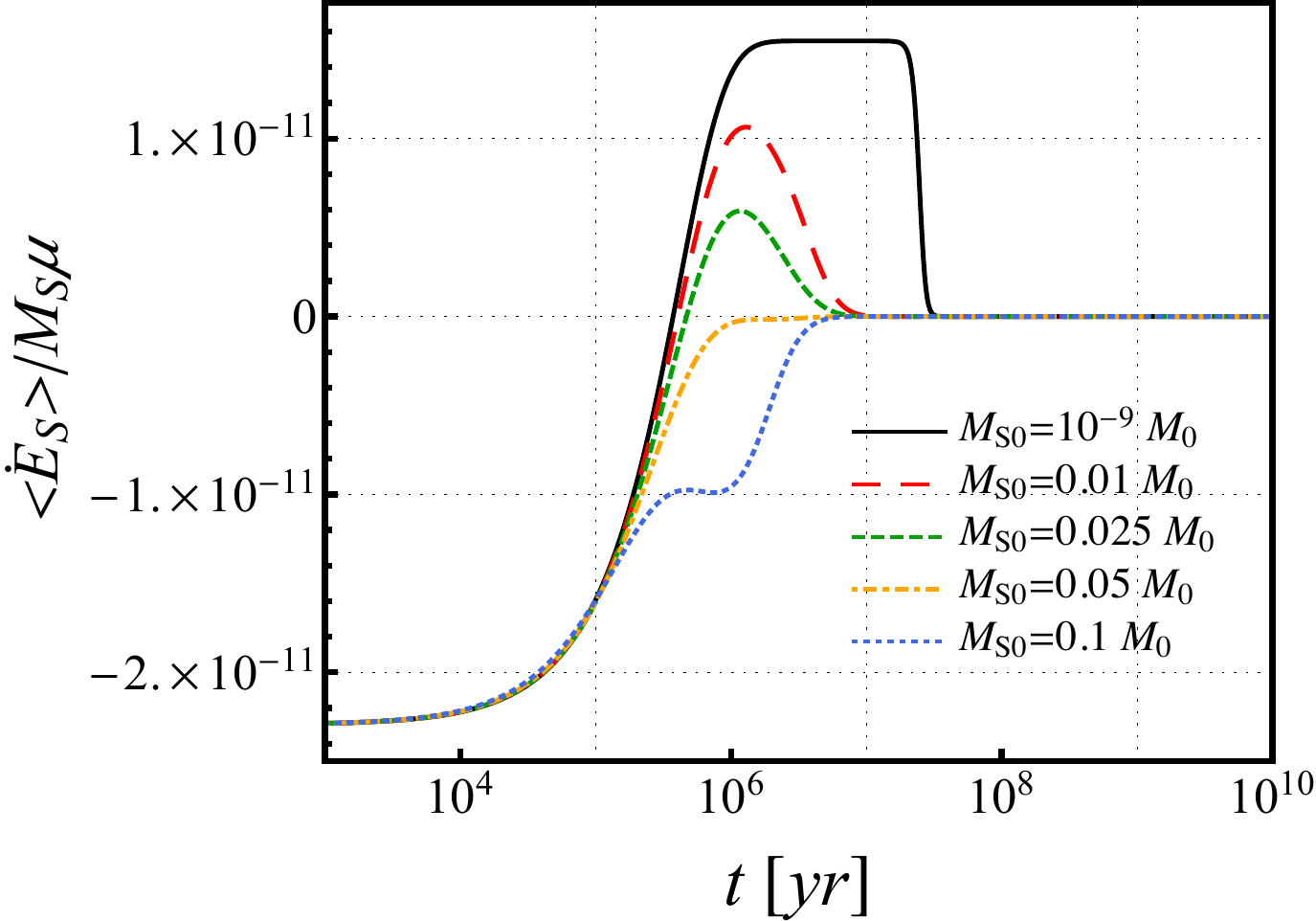}
    \includegraphics[width=.425\textwidth]{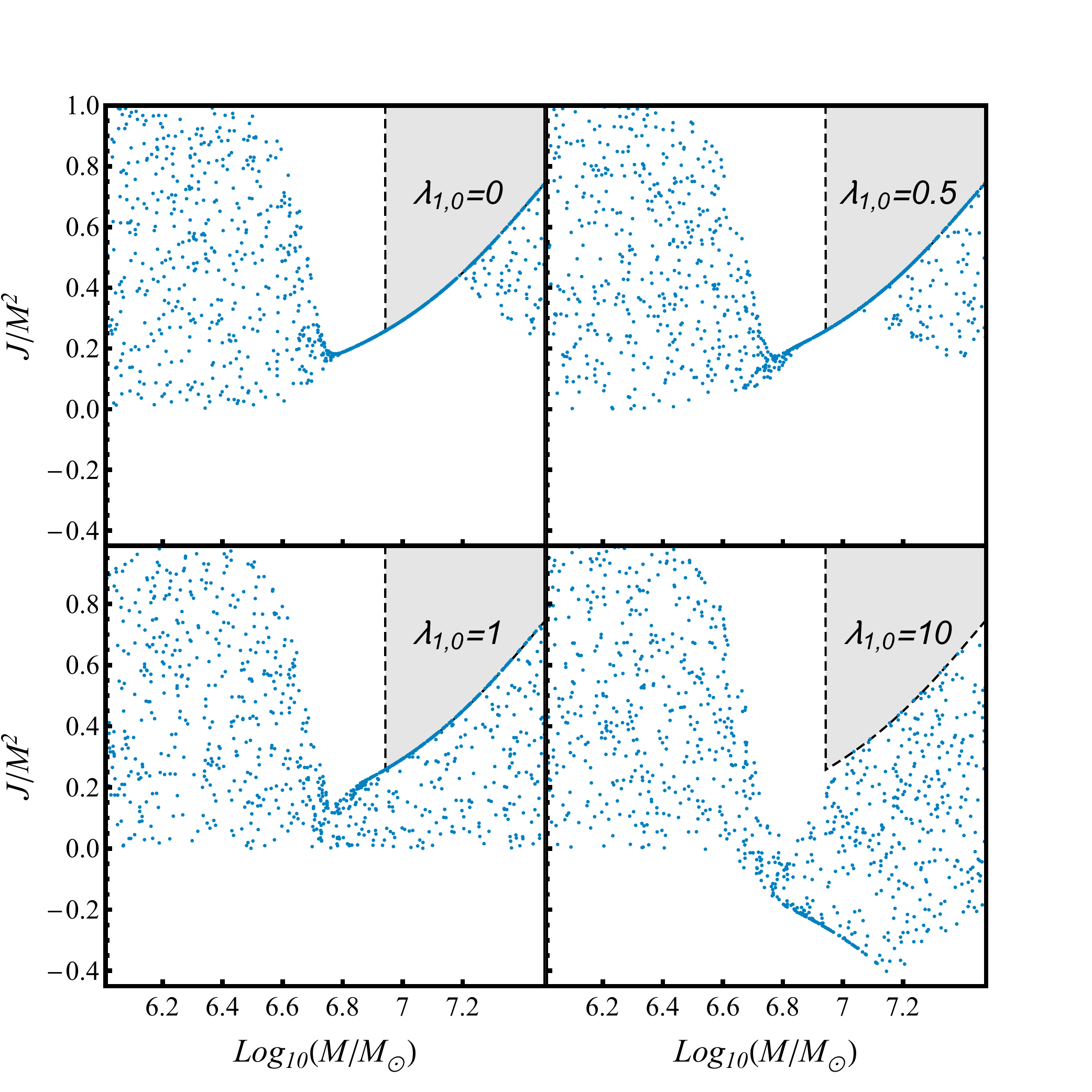}
    \caption{Left: Time evolution of the scalar field energy flux, rescaled by the gravitational coupling $\alpha_G = M\mu_S$, for $M_0 = 10^7 M\odot$, $\chi_0 = 0.8$, $\alpha_{G0} = 0.075$, $\lambda_{1,0} = 1$ and several values of $M_{S0}$ (see main text for details). Right: Black hole Regge plane obtained from the evolution of a scalar field with mass $m_S = 10^{-18}$ eV and containing both superradiant and absorbing modes. We considered scalar cloud initial mass of $M_{S0} = 0.05M_\odot$ and different initial relative amplitudes between the modes $\lambda_{1,0}$.}
    \label{fig:mm_superradiance}
\end{figure}

Superradiant amplification is present when a bosonic field scatters off a rotating black holes with an oscillation frequency $\omega$ satisfies the superradiance condition\cite{Brito:2015oca} $\omega < m \Omega_\mathrm{H}$, where $m$ is the azimuthal index of the field and $\Omega_H$ is the black hole angular velocity. In particular a superradiant instability can naturally develop for massive fields and here we focus on a scalar field of mass $m_S = \mu_S \hbar$ \cite{Detweiler:1980uk}. The phenomenology is ruled by the gravitational coupling $\alpha_G = M \mu_S $ which represents the ratio between the black hole characteristic size and the Compton wavelength of the field $\lambda_C = 1/\mu_S$. the main consequences of this mechanism are two: scalar field develops a macroscopic cloud which will emit gravitational waves at a later stage and the black hole rotational motion is slowed down. Both of these features could be captured by electromagnetic and gravitational observations \cite{Barausse:2020rsu}. The shortest timescale at which this process can occur\cite{Dolan:2007mj} $\tau_\mathrm{S} \sim 50 \left( \frac{M}{M_\odot} \right) \mathrm{s}$ is achieved for $\alpha_\mathrm{G} = 10^{10}\left(M/M_{\odot}\right)\left(\mu_{\mathrm{S}}/\mathrm{eV}\right) \sim \mathcal{O}(1 / 2)$. This implies that for astrophysical black hole masses $\sim\left( 5,10^{10}\right) M_\odot$ one can probe the range $m_B \in \left( 10^{-20}, 10^{-10}\right)$ eV, a regime mostly not accessible by current experiments \cite{Cardoso:2018tly}.

Most studies in the field of superradiant instabilites focus on the case in which the inital data for the scalar field is made of a single superradiant mode. Our goal in this section \cite{Ficarra:2018rfu} is to investigate if the more general case of initial data described by a superposition of a superradiant mode and an absorbing mode can affect the superradiant instability picture. Specifically we added a new parameter to the overall phenomenology which is the initial amplitude ratio $\lambda_{1,0}$ between the modes. Results are presented in Fig.\ref{fig:mm_superradiance} where in the left panel we show the time evolution of the scalar field energy flux at the black hole horizon for an initial black hole mass $M_0 = 10^7 M_\odot$ and dimensionless initial spin $\chi_0 = 0.8$, initial gravitational coupling $\alpha_{G0} = 0.075$, $\lambda_{1,0} = 1$ and several values of the initial energy of the scalar seed $M_{S0}$. In the case of negligible initial mass $M_{S0} = 10^{-9}M_{0}$ the instability follows the single mode phenomenology but for $M_{S0} \gtrsim 5\%M_0$ the superradiant instability is completely quenched. In the right panel of Fig.\ref{fig:mm_superradiance} we present the Regge plane of the black hole (black hole spin vs mass), obtained considering a scalar field with mass $m_S = 10^{-18}$ eV and $M_{S0} = 0.05 M_\odot$ and several values of $\lambda_{1,0}$. The grey shaded area is the Regge gap predicted in the case of a single superradiant initial seed while the blue dots represent multiple modes evolved black holes according to our model. Even though we know superradiance is not present the single mode exclusion region is still present due to absorption of non superradiant modes with opposite angular momentum. Moreover additional forbidden regions appear depending on the value of $\lambda_{1,0}$. This shows that the superradiance mechanism can be turned off in a large region of the parameter space but surprisingly this does not change the fact that the black hole eventually spins down and consequently it does not vary the overall phenomenology. Details were presented in \cite{Ficarra:2018rfu}.

\section{Scalar field dynamics in binary black holes: preliminary results}
In this section we focus on the interplay between scalar fields and binary black holes, which has been recently investigated in the literature \cite{Baumann:2018vus,Baumann:2019ztm,Berti:2019wnn,Ikeda:2020xvt}. In order to complement this studies here we perform numerical relativity simulations of a scalar field during the late inspiral and merger of a black hole binary coalescence using \textsc{Canuda} \cite{witek_helvi_2020_3565475}, a numerical relativity library designed to study fundamental fields in the strong-field regime of gravity. In particular we present preliminary results that will appear soon \cite{Ficarra:2021new}.

\begin{figure}[ht!]
    \centering
    \includegraphics[width=.41\textwidth]{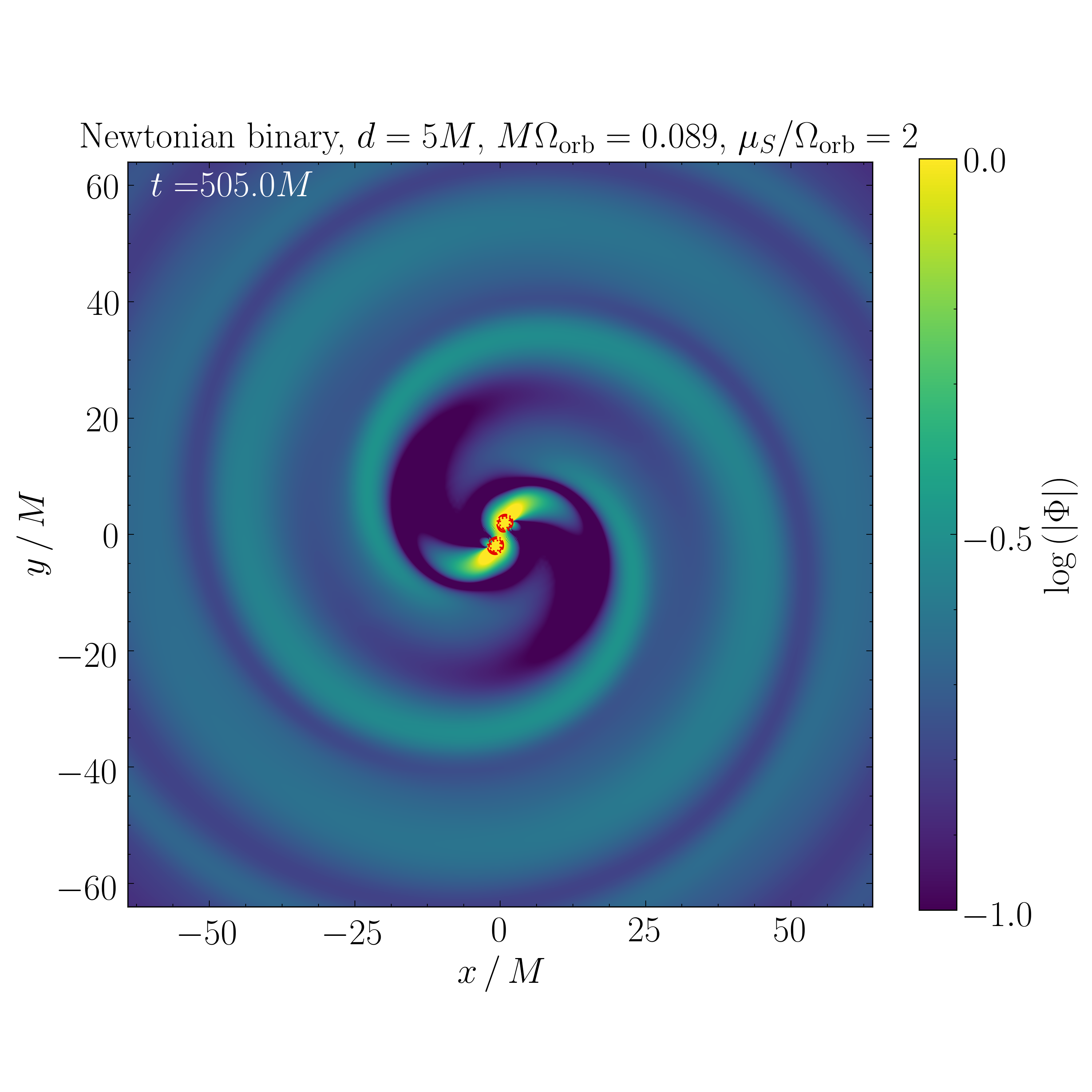}
    \includegraphics[width=.58\textwidth]{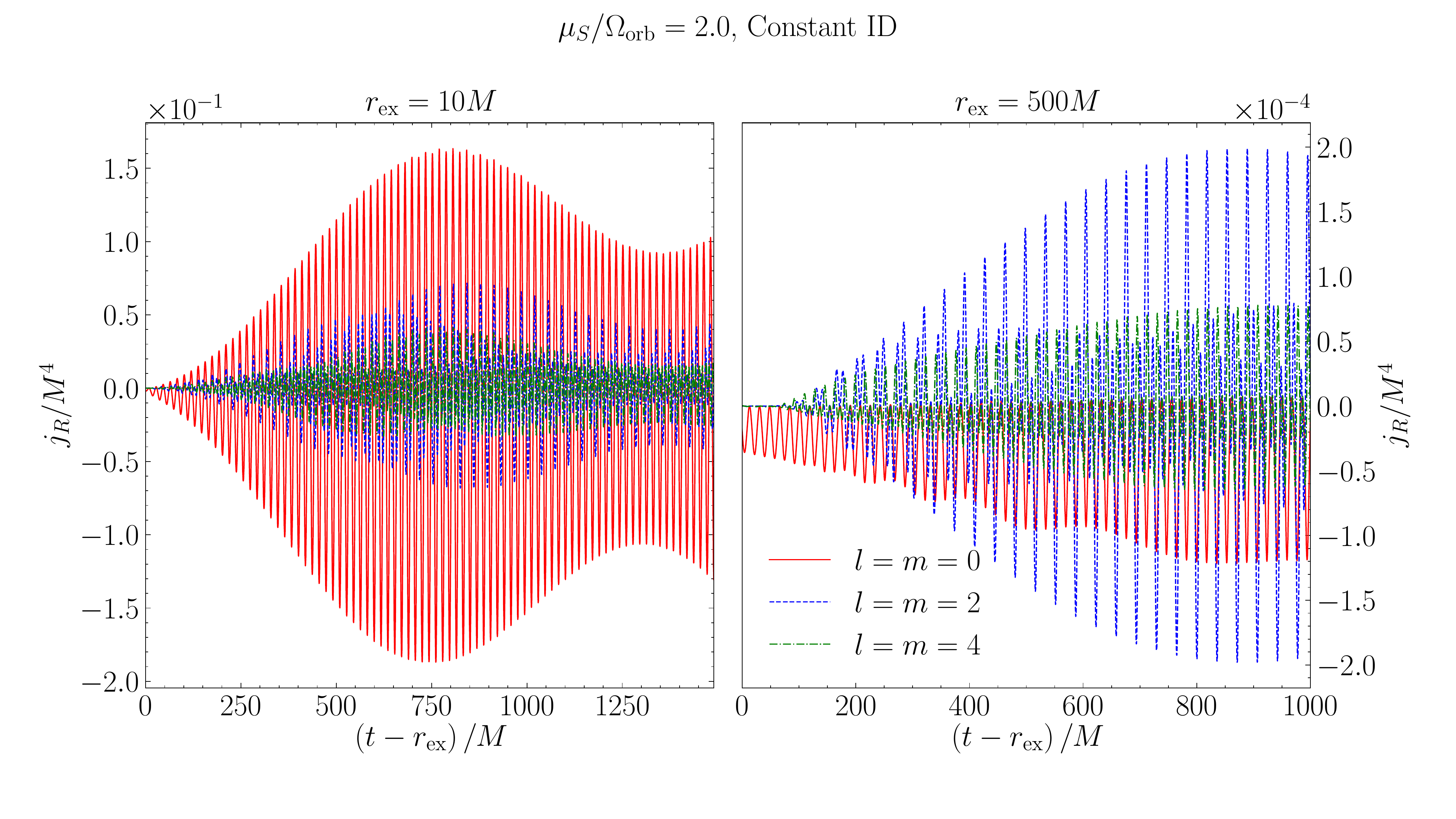}
    \caption{Left: scalar field colour plot in the Newtonian binary orbital plane at $t = 505 M$. This is obtained by considering an orbital separation $d = 5 M$ for which the obrital frequency is $M\Omega_{\mathrm{orb}} = 5 M$. The mass parameter of the field is chosen to be $\mu_S = 2\Omega_{\mathrm{orb}}$. The red circles identify the Newtonian binary companions. Right: scalar field differential energy flux $j_R$ as a function of time at two selected extraction radii $r_{\mathrm{ex}} = 50 M$ and $r_{\mathrm{ex}} = 200 M$, for the same setup as the left panel. Different profiles refer to different multipoles of the scalar field.}
    \label{fig:sf_NBBH}
\end{figure}

The first background spacetime we consider is an equal mass Newtonian binary. The latter is constructed by superposing two Schwarzschild black holes at a fixed orbital separation $d$ with an orbital frequency satisfying Kepler's law $\Omega_{\mathrm{orb}} = \sqrt{M / d^3}$. Results for the time evolution of a scalar field with constant initial profile are depicted in Fig.\ref{fig:sf_NBBH} where we provide a snapshot of the scalar field at a late time in the evolution (left panel). The scalar field in the area near the binary system is characterized by a 10-fold amplification due to accretion of the outer portion of the field on top of the binary. Moreover the rotational motion of the binary excites and propagates away higher multipoles of the field as one can see from the differential energy flux behaviour depicted in the right panel of Fig.\ref{fig:sf_NBBH}.

\begin{figure}[ht!]
    \centering
    \includegraphics[width=.41\textwidth]{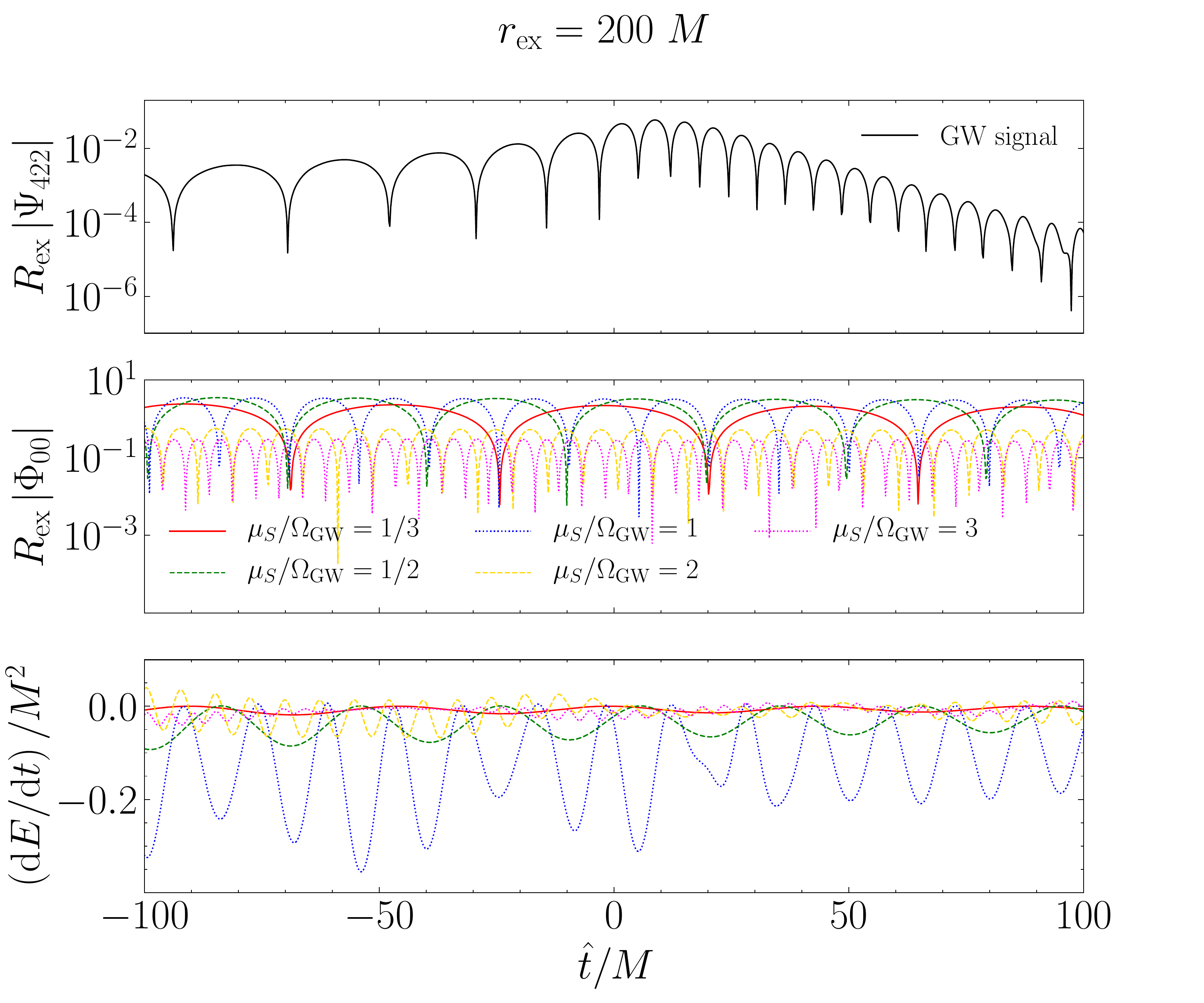}
    \includegraphics[width=.52\textwidth]{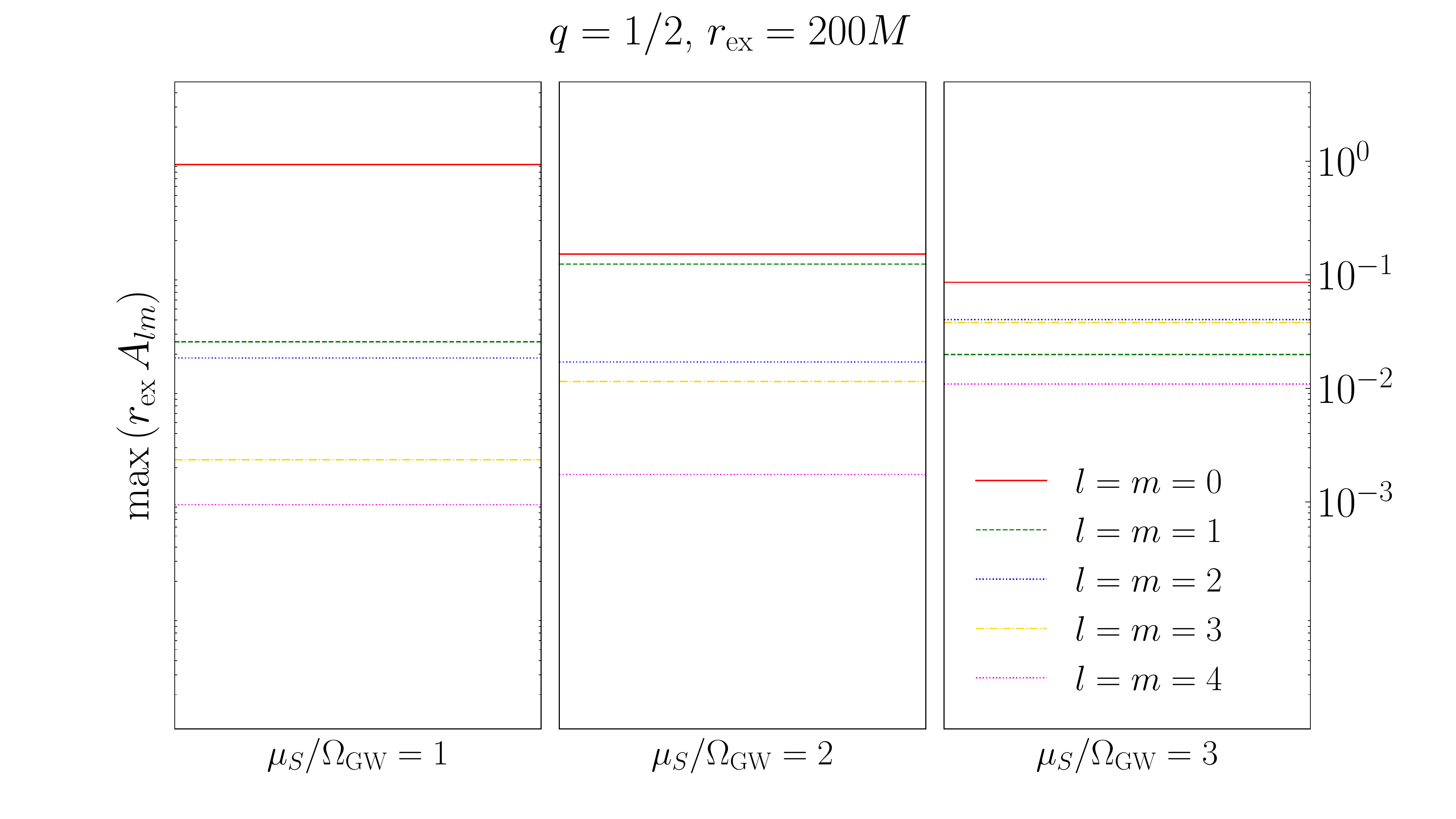}
    \caption{Left: Diagnostic quantities as a function of (rescaled) time $ \hat{t} = \left( t - t_\mathrm{merger} - r_{\mathrm{ex}} \right) / M$ obtained for a fully relativistic black hole binary with mass ratio $q = 1/2$ and an initial orbital separation $d = 10 M$. We show (from top to bottom) the gravitational wave signal $\Psi_{4,22}$, the scalar field monopole $\Phi_{00}$ and total energy flux $dE / dt$ at extraction radius $r_{\mathrm{ex}} = 200 M$. Different profiles are obtained for several values of the ratio between the scalar field mass parameter $\mu_S$ and the gravitational wave frequency a few cycles before merger $\Omega_\mathrm{GW} \sim 0.21 M^{-1}$. Right: scalar field amplitude peak at late times (but before merger) for the same setup as left panel. Three values of $\mu_S / \Omega_{\mathrm{GW}}$ are considered. Different colors show multipolar components of the scalar field.}
    \label{fig:sf_BBH}
\end{figure}

We now move to the description of the time evolution of a scalar field around a fully relativistic black hole binary in the decoupling limit. Results are presented in Fig.\ref{fig:sf_BBH} where we consider a binary black hole of total mass $M$, mass ratio $q = 1/2$, initial separation $d=15M$ and spherically symmetric scalar field initial data. In particular we focus on the late inspiral and merger part of the signal. In the left panel we provide the gravitational waveform $\Psi_{422}$, the scalar field monopole $\Phi_{00}$ and the total scalar energy flux. Also in this case the scalar field is featured by accretion towards the binary, as one can clearly see from the blue dotted line, and this is in agreement with the simpler case of a Newtonian binary. Furthermore the field is characterised by an involved multipolar structure sourced by the rotational motion of the binary as depicted in the right panel of Fig.\ref{fig:sf_BBH}. Interestingly higher multipoles are excited and assume larger amplitudes when the gravitational coupling $\alpha_G$ is of order unity.

\section{Conclusion}
We have investigated the evolution of the black hole superradiant instability against ultralight scalar fields with an initial configuration described by a superposition of modes. In this scenario the instability evolution heavily depends on the energy of the scalar initial seed and on the gravitational coupling. If the seed energy is a few percent of the black hole mass, a black hole surrounded by an overlap of superradiant and absorbing modes with comparable amplitudes might not even experience a superradiant unstable phase. On the other hand, if the seed energy is much smaller than a few percent of the black hole mass the effect of nonsuperradiant modes is negligible. The black hole Regge plane is also affected by the presence of nonsuperradiant modes and its pattern is more involved and additional forbidden regions can appear, depending on the parameters. Moreover we have studied the dynamics of a massive scalar field in a Newtonian binary which magnifies the field due to accretion and propagates higher multipoles through its rotational motion. Finally we analysed a fully relativistic binary black hole which induces a complex multipolar structure in the scalar field.

\section*{Acknowledgments}
G.F. wishes to thank King's College London for financial support and the Royal Society 
for the PhD studentship provided under Research Grant RGF\textbackslash R1\textbackslash 180073.

\begin{footnotesize}
\section*{References}
\bibliography{giuseppeficarra}
\end{footnotesize}

\end{document}